\documentclass[twocolumn,showpacs,preprintnumbers,amsmath,amssymb,nofootinbib]{revtex4}

\usepackage{graphicx,psfrag}
\usepackage{dcolumn}
\usepackage{bm}
\usepackage{multirow}

\begin{document}

\preprint{TU--969}
\preprint{UT-HET-090}

\title{Weight function method for precise determination of top quark mass\\ at Large Hadron Collider}

\author{Sayaka~Kawabata$^a$, Yasuhiro~Shimizu$^a$, Yukinari~Sumino$^a$ and Hiroshi~Yokoya$^b$}
\affiliation{
$^a$Department of Physics, Tohoku University,
Sendai, 980--8578 Japan
\\
$^b$Department of Physics, University of Toyama, 
Toyama, 930--8555 Japan
}%


\begin{abstract}
We propose a new method to measure a theoretically well-defined top quark mass at the LHC.
This method is based on the ``weight function method," which we proposed in our preceding paper.
It requires only lepton energy distribution and is basically independent of the production process of the top quark. 
We perform a simulation analysis of the top quark mass reconstruction with $t\overline{t}$ pair production and lepton+jets decay channel at the leading order.
The estimated statistical error of the top quark mass is about $0.4$\,GeV with an integrated luminosity of $100$\,fb$^{-1}$ at $\sqrt{s}=14$\,TeV.
We also estimate some of the major systematic uncertainties and find that they are under good control.
\end{abstract}

\maketitle

\section{Introduction}
The mass of the top quark is one of the fundamental parameters of the Standard Model (SM) of particle physics. 
It plays a key role among the input parameters in the global SM fit of electroweak precision data through its large contribution to radiative corrections~\cite{Baak:2012kk, Baak:2013fwa}. 
In addition, predictions of models beyond the SM often depend strongly on the value of the top quark mass, e.g. the Higgs boson mass in the minimal supersymmetric SM~\cite{Feng:2013tvd}.
For these reasons, knowing a precise value of the top quark mass is crucial for tests of the SM and models of new physics.
Furthermore, the study of the vacuum stability below the Planck scale within the SM also requires an accurate value of the top quark mass. 
The currently measured values of the masses of the top quark and Higgs boson suggest that the SM vacuum lies close to the border between the stable and meta-stable regions, and the dominant uncertainty in this evaluation comes from the uncertainty of the top quark mass~\cite{Degrassi:2012ry, Buttazzo:2013uya}.

The recent result for the top quark mass obtained by combining direct measurements performed at the Tevatron and Large Hadron Collider (LHC) yields $m_t = 173.34 \pm 0.76$\,GeV~\cite{ATLAS:2014wva}.
These measurements are achieved by using momenta of the top quark decay products including jet momenta and comparing data with Monte Carlo (MC) simulations.
However, since hadronization processes cannot be treated within perturbative QCD, jet momenta depend on modeling of hadronization processes in the MC simulation.
As a result, the top quark mass obtained by such methods is hadronization-model dependent~\cite{Skands:2007zg} and not identical to the pole mass.
It is even difficult to establish the relation between the obtained mass and the pole mass~\cite{Hoang:2008xm}.

Theoretically preferable quark mass definitions are the so-called short-distance masses, in which only short-distance contributions to the self-energy of a quark are renormalized.
A most commonly used one is the mass in the modified-minimal subtraction scheme ($\overline{\rm MS}$ mass), and it is known to exhibit good convergence properties in various perturbative QCD predictions.
So far, the $\overline{\rm MS}$ mass of the top quark has been measured from the $t\overline{t}$ production cross section as $m_t^{\overline{\rm MS}}(m_t^{\overline{\rm MS}})=160.0_{-4.5}^{+5.1}$\,GeV at the Tevatron~\cite{Abazov:2011pta} using the theoretical calculation of the cross section in Ref.~\cite{Moch:2008qy}.
The errors are still large compared to those of the direct measurements mentioned above.
The theoretical errors of the predicted cross section include uncertainties of the parton distribution functions (PDFs), and it would be difficult to reduce these uncertainties significantly in near future.
The PDF uncertainties would limit achievable precision of the $\overline{\rm MS}$ mass to the order of $1$-$2$\,GeV in this method~\cite{Chatrchyan:2013haa, Aad:2014kva}.

Several other methods for measuring the top quark mass at hadron colliders have been proposed~\cite{Chatrchyan:2013boa, Hill:2005zy, Kharchilava:1999yj, Alioli:2013mxa, Biswas:2010sa, Agashe:2012bn}. 
Although these methods complement the conventional direct measurements with different systematic uncertainties, most of them are subject to ambiguities of hadronization models.
Ref.~\cite{Alioli:2013mxa} uses the normalized differential distribution of the $t\overline{t}+1$-jet cross section with respect to the invariant mass of the final state, which has an advantage that ambiguities of hadronization models induce only indirect effects.

In this paper, we propose a new method for a precise measurement of the top quark mass at the LHC.
The method can determine a theoretically well-defined top quark mass.
It is based on the ``weight function method,'' which we proposed in Ref.~\cite{Kawabata:2011gz}.
As proven in Ref.~\cite{Kawabata:2011gz}, using the normalized energy distribution $D(E_\ell)$ of a lepton $\ell$ ($\ell=e {\rm ~or~}\mu$) emitted from the parent particle in the laboratory frame, there exist an infinite number of weight functions $W(E_\ell,m)$ with the following property: the weighted integral,
\begin{equation}
	I(m)=\int dE_\ell D(E_\ell) W(E_\ell,m)\,,
	\label{eq:weightedintegral}
\end{equation}
vanishes when $m$ coincides with the true mass of the parent particle, i.e. $I(m=m^{\rm true})=0$.
This property holds irrespective of the velocity distribution of the parent particle.
The weight functions are constructed with the prediction of the lepton energy distribution in the rest frame of the parent particle, which can be calculated in perturbative QCD.

This method can be used if the parent particle is scalar or unpolarized with respect to the direction of the parent particle boost.
The SM prediction for the longitudinal polarization of the top quark in $t\overline{t}$ pair production at the LHC is $0.003$, generated by the weak interaction~\cite{Bernreuther:2013aga}.\footnote{
One can show using parity invariance that the top quark polarization vector is perpendicular to the top quark boost direction in the QCD production process.}
Since this value is fairly small, our method is applicable (by including small corrections if necessary) to the measurement of the top quark mass, using the lepton in the semileptonic decay of the top quark: $t \rightarrow b\ell\nu$.
Furthermore, since we use only the lepton energy distribution as an observable, this method is (in principle) independent of hadronization models in MC simulations. 
Thus, we can make a direct comparison of the perturbative QCD prediction and experimental data, which enables us to determine the $\overline{\rm MS}$ mass of the top quark.

In the above weight function method, we assume an ideal limit where the narrow-width approximation of the top quark is valid and effects of detector acceptance, event selection cuts and background contributions can be neglected.
There are deviations from the ideal limit, however.
In this first study, we concentrate on the deviations due to experimental aspects, which should be formidable obstacles to achieve an accurate measurement of the top quark mass at hadron colliders.
We perform a simulation analysis of the top quark mass measurement with the weight function method at the leading order (LO), taking account of effects of detector acceptance, event selection cuts and background contributions.
We study top quarks in $t\overline{t}$ production which decay into lepton+jets final states at the LHC.
We show that major experimental problems are estimated to be sufficiently tamed, and therefore, this method can be useful for a precise top quark mass determination.

The paper is organized as follows. 
In Section~\ref{sec:setup}, we explain the setup and basics of our analysis.
In Section~\ref{sec:cuts}, we examine effects of event selection cuts.
We show the results of top mass reconstruction in Section~\ref{sec:results}.
We discuss possible problems and solutions in Section~\ref{sec:discussion}.
Conclusion is given in Section~\ref{sec:conclusion}.

\section{Setup and basics of analysis}
\label{sec:setup}
We consider for the signal process, $t\overline{t}$ productions and their subsequent decays into a muon plus jets at the LHC:
\begin{equation}
	pp\rightarrow t\overline{t} +X \rightarrow \mu^{\pm}+{\rm jets}+p_T^{\rm miss}\,.
\end{equation}
For the background events, other $t\overline{t}$, $W+$jets, $Wb\overline{b}+$jets and single-top processes are considered.
Other $t\overline{t}$ events include all the decay channels of $t\overline{t}$ except the above signal decay channel.
In particular, the other lepton+jets channel $e+$jets, as well as the $\mu+$jets final state, where $\mu$ is produced in tau lepton decays, are regarded as background events.
The $W+$jets and $Wb\overline{b}+$jets events are generated by using the matrix elements of $W+\{0,1,2,3,4\}$-jets and $Wb\overline{b}+\{0,1,2\}$-jets processes merged with parton-shower, respectively.
The single-top background includes contributions from the $t$-channel, $s$-channel and $Wt$ associated production processes.

Both signal and background events are generated at LO using MadGraph/MadEvents~\cite{Maltoni:2002qb} with $\sqrt{s}=14$\,TeV, and they are passed to PYTHIA~\cite{Sjostrand:2006za} subsequently. 
We use the parton distribution function CTEQ6L~\cite{Pumplin:2002vw}.
All these events are passed to the fast detector simulator PGS~\cite{PGS}.
Pileup events are not considered.
We generate $8\times10^{6}$ events for the signal process.

\begin{figure}[t]
\includegraphics[width=6cm]{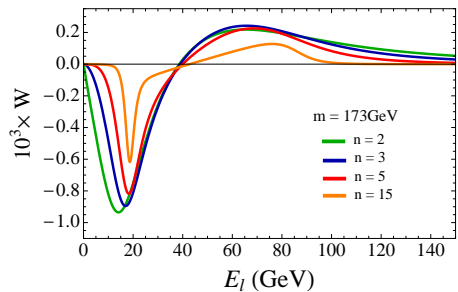}
\vspace*{-2mm}
\caption{\small
Weight functions $W(E_\ell,m)$ used in the analysis with $m=173$\,GeV, corresponding to $n=2,\,3,\,5$ and $15$ in Eq.~(\ref{eq:oddfunc}).
\label{fig:weightfunc}
}
\vspace{-3mm}
\end{figure}

The explicit form of a weight function is given by~\cite{Kawabata:2011gz}
\begin{equation}
	W(E_\ell,m)=\!\int \!dE \left.\mathcal{D}_0(E;m)\frac{1}{EE_\ell} \,({\rm odd~fn.~of~}\rho)\right|_{e^{\rho}=E_\ell/E}\,,
	\label{eq:weightfunc}
\end{equation}
where $\mathcal{D}_0(E;m)$ is the normalized lepton energy distribution in the rest frame of the parent top quark with the mass $m$.
We can choose an arbitrary odd function of $\rho$ in the bracket of the integrand.
At LO, $\mathcal{D}_0(E;m)$ is given by
\begin{eqnarray}
	\mathcal{D}_0(E;m)&\propto&E\left\{ \frac{m}{2}\left( 1-\frac{m_b^2}{m^2}\right)-E\right\}\nonumber\\
	&&\times\left\{\arctan \!\left(\frac{m_W}{\Gamma_W}\right)
	-\arctan\! \left(\frac{m_W^2-\mu^2_{\rm max}}{m_W \Gamma_W}\right)\right\}\nonumber\\
	&&\times\,\,\theta(\,0< E<E_{\rm max}\,)\,,
\end{eqnarray}
with
\begin{eqnarray}
	E_{\rm max}&\equiv&\frac{m^2-m_b^2}{2m}\,,\\
	\mu^2_{\rm max}&\equiv&\frac{2E(m^2-m_b^2-2mE)}{m-2E}\,,
\end{eqnarray}
where $m_b$, $m_W$, and $\Gamma_W$ represent the masses of the bottom quark, $W$ boson, and the width of the $W$ boson, respectively.
We choose for the odd function of $\rho$ in Eq.~(\ref{eq:weightfunc}) as
\begin{equation}
	({\rm odd~fn.~of~}\rho)\,=\,n\tanh (n\rho)/\!\cosh(n\rho)\,,
	\label{eq:oddfunc}
\end{equation}
with $n=2,\,3,\,5$ and $15$.
Fig.~\ref{fig:weightfunc} shows the weight functions used in the following analysis obtained from Eq.~(\ref{eq:weightfunc})-(\ref{eq:oddfunc}) with $m=173$\,GeV.
Because of their small weights around $E_\ell=0$, where effects of the lepton cuts are large (see next section), we can expect that these weight functions are less sensitive to the cuts, compared to weight functions which have non-zero values around $E_\ell=0$.

\begin{figure}[t]
\includegraphics[width=6cm]{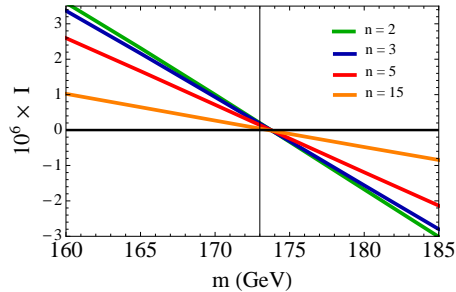}
\vspace*{-2mm}
\caption{\small
Weighted integrals $I(m)$ defined by Eq.~(\ref{eq:weightedintegral}) with the parton-level lepton distribution and the weight functions corresponding to $n=2,\,3,\,5$ and $15$. The input value of the top quark mass is $173$\,GeV.
\label{fig:ImParton}
}
\vspace{-3mm}
\end{figure}

To check validity of the weight function method, we first examine the method at the parton level.
We use for $D(E_\ell)$ in Eq.~(\ref{eq:weightedintegral}) the lepton energy distribution of the signal MC events at the parton level.
Computing the weighted integrals $I(m)$ [Eq.~(\ref{eq:weightedintegral})], we obtain Fig.~\ref{fig:ImParton}, corresponding to the weight functions of Fig.~\ref{fig:weightfunc}. 
The input value of the top quark mass in the MC events is $173$\,GeV. 
On the other hand, the zeros of $I(m)$ are located around $173.7$\,GeV.
Estimated statistical errors of the MC simulation are around $0.4$\,GeV, and due to the effect of the top width, the mean value of the top invariant mass at the parton level is shifted from the input top quark mass by $+0.34$\,GeV in our analysis.\footnote{
We use the Breit-Wigner distribution in MC with a cut-off at $m_t\pm \Gamma_t\times 50$, where $m_t$ and $\Gamma_t$ are the mass and width of the top quark.
The shift $+0.34$\,GeV due to the effect of the top width is a systematic bias to be corrected.
}
Therefore, we confirm that our method works within the MC statistical errors.

\section{Effects of event selection cuts}
\label{sec:cuts}
In this section we examine effects of various event selection cuts.
We make several assumptions and take specific analysis methods.
Some of these assumptions and analysis methods need to be examined carefully, since they can be sources of systematic uncertainties.
We provide further discussion on these points in Section~\ref{sec:discussion}.

\begin{figure}[t]
\includegraphics[width=6cm]{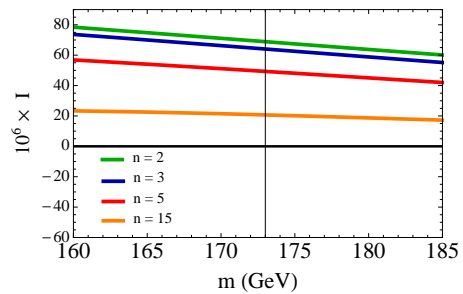}
\vspace*{-2mm}
\caption{\small
Weighted integrals $I(m)$ with the MC events after the lepton cuts for the weight functions corresponding to $n=2,\,3,\,5$ and $15$. The input value of the top quark mass is $173$\,GeV.
\label{fig:ImAftLepCuts}
}
\vspace{-3mm}
\end{figure}

In  real experiments, detector effects, event selection cuts and backgrounds deform the lepton energy distribution.
The major effect is caused by the lepton cuts:
\begin{equation}
	p_T(\mu)>20{\rm \,GeV},~~|\eta(\mu)|<2.4\,,
	\label{eq:leptoncuts}
\end{equation}
where $p_T(\mu)$ and $\eta(\mu)$ are the transverse momentum and pseudo-rapidity of a muon, respectively.
For the values of the above $p_T(\mu)$ and $\eta(\mu)$ cuts, we refer to the LHC $p_T$ trigger~\cite{Aad:2009wy, Ball:2007zza} and the PGS default value of $\eta$ cut.
The lepton cuts reduce mainly the low-energy part of the lepton distribution.
This results in large shifts of the weighted integrals $I(m)$, as shown in Fig.~\ref{fig:ImAftLepCuts}.
Because the weight functions are negative for small $E_\ell$, where the lepton distribution is largely reduced, the weighted integrals shift in the positive direction.
The zeros of $I(m)$ are significantly displaced from the input top mass due to these shifts.

We solve this problem by compensating for the loss caused by the lepton cuts, using MC events which satisfy $p_T(\mu)<20$\,GeV or $|\eta(\mu)|>2.4$.
This is because (1) experimental effects are well understood concerning leptons, so that the estimates of MC simulations are accurate for the lepton distribution, and
(2) the weight function method utilizes the fact that the angular distribution of the lepton in the rest frame of the top quark is flat.\footnote{
To be precise, the lepton $\cos \theta_\ell$ distribution is (almost) flat, where $\theta_\ell$ is measured from the boost direction of the top quark in the rest frame of the top quark.
}
When this condition holds, the zero of $I(m)$ is independent of the velocity distribution of the top quark, owing to which we do not need information on the velocity distribution.
Therefore, in order to make maximum use of the advantages of this method, we recover the flat angular distribution of the lepton and return the zero of $I(m)$ to the right place.
The normalization of the compensated events is determined such that the $p_T(\mu)$ distribution of the data and compensated events are connected smoothly.
We can check validity of the compensated events partly, using the di-leptonic channel, whose lepton $p_T$ cut can be looser than the lepton$+$jets channel.
We evaluate part of uncertainties in the compensated events by varying the factorization scale and PDF in the MC in Section~\ref{sec:results}.

The effects which cause differences between the lepton momenta at the parton level and detector level such as the effects of lepton isolation and photon emissions, also deform the lepton energy distribution.
Since these effects are also well understood, it should be possible in principle to estimate them and restore the parton-level lepton distributions.
We assume that the distributions are restored not event by event but for their entire distributions.
In this simulation analysis, we suppose for simplicity that they can be estimated and restored completely for signal events and just use the parton-level leptons.
We use the detector-level information of each event for the lepton momenta of the background events and jet momenta.

The criterion for validity of the weight function method is whether we can restore the original shapes of the following two distributions by unfolding detector effects and effects of cuts and backgrounds: the lepton energy and angular distributions in the rest frame of the top quark.
(Typical required accuracies are a few percent level.)
Cuts concerning missing transverse momentum $p_T^{\rm miss}$ affect mainly the former distribution through the following reason.
In the rest frame of the top quark, the energy of a neutrino is uniquely determined from the energy of its paired lepton neglecting the effects of the $W$ off-shellness and at LO.
In this frame, high-energy leptons corresponds to low-energy neutrinos.
This tendency remains after including the effects of boosts with the top quark velocity distribution. 
Since the angular distribution of the neutrino in the rest frame of the top quark is flat just like the lepton, neutrinos with an energy $E_\nu^{\rm \,rest}$ in the top quark rest frame have a distribution in the laboratory frame with its peak in the same position as the rest frame energy $E_\nu^{\rm \,rest}$, that is, many neutrinos tend to keep the same energy after the boosts~\cite{Agashe:2012bn}.
Thus, a high-energy lepton in the top quark rest frame still corresponds to low-energy part of neutrino after the boosts, and for example, dropping events with small $p_T^{\rm miss}$ corresponds to losing mainly a high-energy part of lepton distribution in the top quark rest frame.
For this reason, tight $p_T^{\rm miss}$ cuts severely deform the lepton distribution, and thus, we do not apply cuts concerning $p_T^{\rm miss}$ in this analysis.
In order to apply this method in real experiments, $p_T^{\rm miss}$ cuts need to be relaxed compared to those used in current major analyses.
We confirmed that loose cuts concerning $p_T^{\rm miss}$, for example, $p_T^{\rm miss}>4$\,GeV and $E_\ell+p_T^{\rm miss}>33$\,GeV, are also applicable keeping the above criterion.

Cuts using $b$-tagging are highly effective to reduce the $W+$jets background.
However, $p_T$ and $\eta$ dependence of the $b$-tagging efficiency causes a bias to the lepton distribution, especially to the angular distribution.
One way to cope with this effect is using the estimate of the $b$-tagging efficiency $\epsilon_b(p_T, \eta)$, and multiplying the lepton energy distribution by $\epsilon_b^{-1}(p_T,\eta)$.
Although this is a simple and realistic way, it is affected by the experimental accuracy of $\epsilon_b(p_T, \eta)$ and depends on the details of the actual experimental conditions.
In this simulation analysis, we apply an alternative way which causes nearly equivalent effects but more conservative results.
We apply $b$-tagging with an efficiency independent of $p_T$ and $\eta$ to taggable $p_T>15$\,GeV and $\eta<2.5$ $b$-jets, adopting the lowest efficiency in the taggable region.
We refer the value of the ATLAS experiment for the $b$-jets taggable region~\cite{Aad:2009wy}.
This flat efficiency would be attainable in experiments in principle.

To simulate the flat $b$-tagging efficiency, we assume that the expectation value of the number of $b$-jets within the above taggable region is $N_b(n/N)$ for events with $N_b$ bottom quarks, where $n$ and $N$ are the numbers of jets in the taggable region and all region, respectively.\footnote{
This assumption is not true by $10$-$20$\,\%, and as a result, we underestimate the efficiency.
}
We use the following values for the efficiencies in the region $p_T>15$\,GeV and $|\eta|<2.5$,
\begin{eqnarray}
	&b{\rm \mathchar`-tagging~efficiency:}&40\,\%\,,\nonumber\\
	&{\rm mis\mathchar`-tagging~rate~for~light~jets:}&0.5\,\%\,,\nonumber
\end{eqnarray}
referring to the lowest $b$-tagging efficiency in Ref.~\cite{Aad:2009wy}.
We choose $b$-tagged events randomly according to the probability derived from the above assumption.

Considering the effects of cuts as explained above, we impose the following event selection cuts to the MC events:
\begin{itemize}
\item One muon with $p_T>20$\,GeV and $|\eta|<2.4$\,.
\item At least four jets.
\item At least one $b$-tagged jet with the $b$-tagging efficiency $0.4$ independent of $p_T$ and $\eta$ in the region $p_T>15$\,GeV and $|\eta|<2.5$\,.
\item $p_T(j_1)>55,~p_T(j_2)>25,~p_T(j_3)>15,~p_T(j_4)>8$\,GeV,
\end{itemize}
where $p_T(j_i)$ is the transverse momentum of the jet with the $i$-th largest $p_T$.
Here, we do not regard the hadronically-decaying tau lepton as a jet.

The LO cross sections after all the event selection cuts are summarized in Table~\ref{tab:crosssection}.\footnote{We do not apply corrections due to K-factors.}
They are evaluated using events at the detector level.

\begin{table}[t]
	\centering
	\begin{tabular}{c|cccc}
		\hline
		Signal~(pb)&\multicolumn{4}{c}{Background~(pb)}\\
		~$m_t=173$\,GeV~&~Other $t\overline{t}$~~&~$W+$jets~&~~$Wb\overline{b}+$jets~~&Single top\\
		\hline 
		 $22.4$ & $5.7$ & $1.8$ & $1.8$ & $1.3$\\
		\hline
	\end{tabular}
	\caption{\label{tab:crosssection} Cross sections after all the cuts.}
\end{table}

\section{Results of top mass reconstruction}
\label{sec:results}

Our strategy is as follows: after all the cuts are applied, background contributions to lepton distributions are estimated and subtracted.
In addition, the effects which cause differences between lepton momenta at the parton level and detector level are estimated, and the parton-level lepton distributions are (assumed to be) restored for the signal events.
After these procedures, the following top quark mass reconstruction is performed.
Let us first perform the analysis with only the signal events after the cuts.
Later we consider effects of background events.

For the events after all the cuts, we compensate the loss caused by the lepton cuts using MC events at the parton level, as mentioned in the previous section.
To determine the normalization of the compensated events, we perform a $\chi^2$-fit to the $p_T(\ell)$ distribution so that the $p_T(\ell)$ distributions of the data and compensated events are connected smoothly around $p_T(\ell)=20$\,GeV.
The setup of the $\chi^2$-fit is as follows.
We choose the range of $p_T(\ell)$ for the fit to be $[17,30]$\,GeV with the binwidth of $1$\,GeV.
We take a narrower range below $20$\,GeV because the number of the compensated events in each bin below $20$\,GeV is large compared to that of the data (which is above $20$\,GeV), and it is desirable that the fit does not depend strongly on the MC part, but on the data part.
The numbers of events in these bins are fitted to a quartic function, together with the normalization of the compensated events taken as a free parameter.

The validity of the fit was partly checked using $10^3$ toy MC samples generated for both the compensated part and data part.
We used simple functions fitted to the lepton $p_T$ distributions as ``true" distributions for simplicity.
We repeated the $\chi^2$-fit for all of the samples.
In addition, using the toy MC results, we estimate a statistical error of the normalization in this simulation analysis. 

Ideally, a large number of the compensated events can result in a small MC statistical error originating from the compensated part.
In this analysis, however, we generate about $(17$-$18)\times10^5$ events for the compensated part, which correspond to the amount of data with about $100$\,fb$^{-1}$, and the MC statistical error from the compensated part is not negligible.
On the other hand, in analyses of real experiments, we expect that $10$-$100$ times the size of real data can be generated for the compensated part.
In this case, MC statistical errors from the compensated part can be insignificant.

\begin{figure}[t]
\includegraphics[width=6cm]{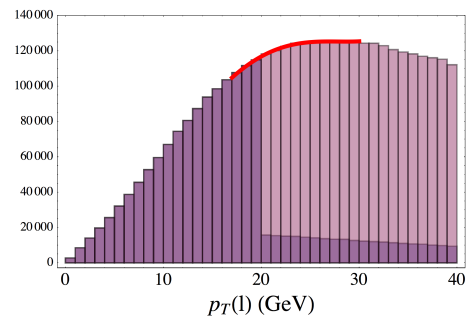}
\vspace*{-2mm}
\caption{\small
Sum of the lepton $p_T$ distributions of the compensated events normalized using a $\chi^2$-fit (dark purple) and the events after all the cuts (light pink). The fitted function is also shown as a red line.
\label{fig:fitting}
}
\vspace{-3mm}
\end{figure}

To illustrate this fitting method, Fig.~\ref{fig:fitting} shows the sum of the lepton $p_T$ distributions of the compensated events normalized by the above method (dark purple) and the events after all the cuts (light pink).
The fitted function is also shown as a red line.
The value of the top quark mass of the compensated MC events need to be assumed and we call this $m_t^{\rm c}$.
Both input value of the top quark mass and $m_t^{\rm c}$ are taken to be $173$\,GeV in this figure.

\begin{figure}[t]
\includegraphics[width=6cm]{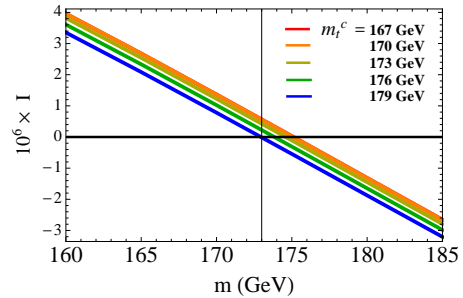}
\vspace*{-2mm}
\caption{\small
Weighted integrals $I(m)$ with various $m_t^{\rm c}$ after all the cuts. The weight function used corresponds to $n=2$ in Eq.~(\ref{eq:oddfunc}). The input value of the top quark mass is $173$\,GeV.
\label{fig:ImWithmtMCAftCuts}
}
\vspace{-3mm}
\end{figure}

We construct the weighted integrals $I(m)$, using the lepton energy distributions of the events after all the cuts with the input top quark mass $173$\,GeV and the compensated events with various $m_t^{\rm c}$.
Fig.~\ref{fig:ImWithmtMCAftCuts} shows the weighted integrals $I(m)$.
In this plot, we use the weight function corresponding to $n=2$.
Although $m_t^{\rm c}$ vary from $167$ to $179$\,GeV, the variation of the zero of $I(m)$ is much less.

\begin{figure}[t]
\includegraphics[width=6cm]{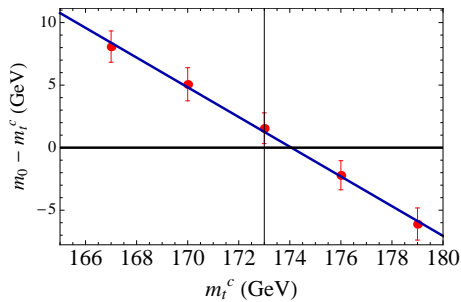}
\vspace*{-2mm}
\caption{\small
The zero of $I(m)$, $m_0$, minus $m_t^{\rm c}$ as a function of $m_t^{\rm c}$ (red points). The error bars correspond to the estimated statistical errors of the MC simulation. The weight function used corresponds to $n=2$. The input value of the top quark mass is $173$\,GeV. The blue line shows the linear function fitted to the red data points.
\label{fig:m0-mtCAftCuts}
}
\vspace{-3mm}
\end{figure}

From the zeros of $I(m)$, we can reconstruct the top quark mass in the following manner:
if $m_t^{\rm c}$ is equal to the input mass, the zero of $I(m)$ (denoted as $m_0$) should be $m_t^{\rm c}$.
In contrast, if $m_t^{\rm c}$ is different from the input mass, there is no guarantee that $m_0$ equals $m_t^{\rm c}$ and it is expected to be different from $m_t^{\rm c}$.
Therefore, we obtain the value of $m_t^{\rm c}$ where $m_0$ coincides with $m_t^{\rm c}$ as the reconstructed mass: $m_t^{\rm rec}=m_t^{\rm c}\,(m_0=m_t^{\rm c})$.
Fig.~\ref{fig:m0-mtCAftCuts} shows $m_0 - m_t^{\rm c}$ as a function of $m_t^{\rm c}$.
The fitted linear function is also shown.
The zero of the fitted function is at $174.1$\,GeV.
The error bars correspond to the estimated statistical errors of the MC simulation obtained from evaluation of the normalization of the compensated part.
They include errors from both the compensated part and data part.
Consequently the MC statistical error for the reconstructed top quark mass is $+1.0/\!\!-\!1.1$\,GeV for the weight function of $n=2$, and the shift expected from the effect of the top width is $+0.34$\,GeV.
Thus, the size of the shift $+1.1$\,GeV from the input top quark mass is consistent with their effects.

\begin{figure}[t]
\includegraphics[width=6cm]{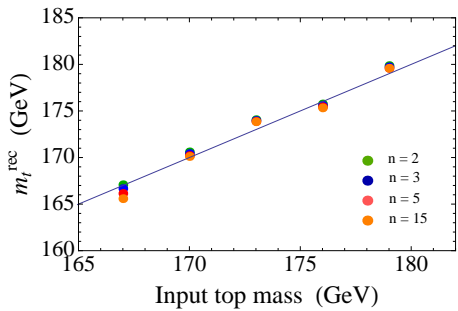}
\vspace*{-2mm}
\caption{\small
Reconstructed top quark mass as a function of the input mass. The weight functions used correspond to $n=2,\,3,\,5$ and $15$. The blue line shows the line where the reconstructed mass is equal to the input mass.
\label{fig:MtMeasAftCuts}
}
\end{figure}

We perform the same top mass reconstruction as stated above for various input values of the top quark mass and various weight functions.
The obtained results are shown in Fig.~\ref{fig:MtMeasAftCuts}.
The vertical axis is the reconstructed top quark mass obtained with this method and the horizontal axis is the input top quark mass of the events.
The blue line shows the line where the reconstructed mass is equal to the input mass, i.e. the ideal measurement.
The values of the results for the weight function corresponding to $n=2$ are also shown in Table~\ref{tab:measuredmass}.
Considering the effects of the top width on the measured masses, whose sizes are $+0.3$ to $+0.4$\,GeV depending on the top quark mass, and the MC statistical errors, the reconstructed masses are consistent with the input masses.

\begin{table}[t]
	\centering
	\begin{tabular}{c|ccccc}
		\hline
		Input top mass(GeV) & ~~$167$~~ & ~~$170$~~ & ~~$173$~~ & ~~$176$~~ & ~~$179$~~\\
		\hline 
		$m_t^{\rm rec}$(GeV) & $167.1$ &$170.6$& $174.1$ & $175.7$ & $179.9$\\
		\hline
	\end{tabular}
	\caption{\label{tab:measuredmass} Reconstructed top quark mass as a function of the input mass. The weight function used corresponds to $n=2$.}
\end{table}

We estimate uncertainties from several sources in the top mass reconstruction.
Besides signal and background statistical errors, we estimate uncertainties originating from the dependences on the factorization scale, PDF and jet energy scale (JES).
Since we use MC simulations for the compensated events in this method, the factorization scale and PDF uncertainties in the MC can be serious.
In addition, the JES uncertainty is one of the largest uncertainties in the conventional direct measurements of the top quark mass~\cite{ATLAS:2014wva}.

Table~\ref{tab:uncertainties} shows the results of the estimates.
The input value of the top quark mass in these estimates is $173$\,GeV.
The signal statistical errors are estimated as follows:
we divide the generated events into $15$, $20$, $50$ and $100$ subgroups of equal sizes and perform the same top mass reconstruction as explained in this section for each sample.
Results of the fits to determine the normalization of the compensated events depend on the number of events in each sample.
Thus, the statistical errors obtained from the standard deviations of reconstructed mass distributions depend on the number of the division.
We extrapolate statistical errors at the number of events for $100$\,fb$^{-1}$ from the results of these subgroups.
Assuming that the errors of the electron mode is the same as the muon mode, we estimate the statistical error of the sum of lepton+jets events, i.e. the combination of the muon and electron modes.
The uncertainties from the factorization scale dependence of the signal events are estimated by changing the scale in the compensated MC events by $1/2$ and $2$.
The errors from the PDF uncertainties are estimated using different sets of PDFs, MSTW2008~\cite{Martin:2009iq} and NNPDF2.1~\cite{Ball:2011uy}, for the compensated events.
The uncertainties associated with JES are estimated by varying the $p_T$ of all jets in events by $\pm10$\,\% before the event selection cuts.\footnote{
Since in the event selection cuts we choose $b$-tagged events randomly according to their probabilities, this cut involves statistical error.
In order to obtain the uncertainties purely from JES without the statistical error, the requirement of a $b$-tagging is excluded from the event selection cuts in this estimation.
}

The background statistical errors are estimated as follows.
Since other $t\overline{t}$ events are the dominant source of backgrounds after the cuts, we focus on their effects.
We obtain about $6\times 10^4$ MC events after the cuts.
The lepton energy-$p_T$ distribution is modeled by a simple function imitating these events.
Using this function, we generate a few tens MC samples of the lepton distribution for background events.
We add to the signal events each of these samples and subtract an ``estimated" lepton distribution, which is constructed from the modeling function.
With these sets of signal-plus-background-errors, we simulate top mass reconstruction in our method and compute the standard deviations of the reconstructed top quark mass.\footnote{
The numbers of signal and background events in each set correspond to about $10$\,fb$^{-1}$ integrated luminosity.
Only in this analysis of the background effects we use a smaller number of signal MC events (as compared to other analyses in this section) to save analysis time.
}
After rescaling according to square-root of the number of events, we obtain for $100$\,fb$^{-1}$ integrated luminosity the estimates listed in Table~\ref{tab:uncertainties}.

In addition, we estimate shifts of the reconstructed top mass in the case that we mistake by $5$\,\% the normalization of the estimated background contribution, which we subtract from the measured lepton distribution.\footnote{
Note that the major background is other $t\overline{t}$ events whose production process is the same as that of the signal, and therefore, the $5$\,\% mistake of the {\it ratio} of the normalization is rather conservative.}
If we do this naively, we find that the shifts are about $1.5$-$2$\,GeV for $n=2,\,3,\,5$ and $15$.
However, we also find that the size of the shifts depends strongly on the setup of the fitting for the determination of the normalization of the compensated events.
This is because the shapes of $p_T(\ell)$ and $E_{\ell}$ distributions are different between the signal and background.
Compared to the signal, these distributions of the other $t\overline{t}$ background show sharp declines in high-energy regions.
If we take into account these general features to choose the range and the function for the fit, the shifts can be kept as moderate as $0.6$-$0.8$\,GeV (depending on $n$).

\begin{table}[t]
	\centering
	\begin{tabular}{l|cccc}
		&\multicolumn{4}{c}{n}\\
		&~~~~~$2$~~~~~&~~~~~$3$~~~~~&~~~~~$5$~~~~~&~~~~~$15$~~~~~\\
		\hline \hline
		Signal stat. error & $0.4$& $0.5$ & $0.5$ & $0.6$\\ \hline
		\multirow{2}{*}{Fac. scale (signal)} & $+1.5$& $+1.4$ & $+1.4$ & $+1.4$\\
		 & $-1.4$& $-1.3$ & $-1.2$ & $-1.2$\\ \hline
		PDF (signal) & $0.6$ & $0.8$ & $1.1$ & $1.4$\\ \hline
		\multirow{2}{*}{Jet energy scale (signal)} & $+0.2$ & $+0.3$ & $+0.4$ & $+0.5$\\ 
		& $-0.0$ & $-0.2$ & $-0.4$ & $-0.5$\\ \hline
		Background stat. error & $0.4$ & $0.4$ & $0.4$ & $0.4$\\
		\hline
	\end{tabular}
	\caption{\label{tab:uncertainties} Estimates of uncertainties in GeV from several sources in the top mass reconstruction. The weight functions used correspond to $n=2,\,3,\,5$ and $15$. The input value of the top quark mass used in the estimates is $173$\,GeV. The signal statistical errors correspond to those with an integrated luminosity of $100$\,fb$^{-1}$ and for the sum of the lepton($e,\mu$)+jets events. The background statistical errors are also for $100$\,fb$^{-1}$.}
\end{table}

One can see in Table~\ref{tab:uncertainties} that the uncertainties from the factorization scale dependence dominate.
The JES uncertainties are relatively small, reflecting the characteristics of our method which uses solely the lepton distribution.
Combining the uncertainties in Table~\ref{tab:uncertainties}, the total uncertainty amounts to about $1.7$\,GeV for $n=2$.

\section{Discussion}
\label{sec:discussion}

The important point in our method is how accurately the parton-level signal distributions of leptons can be restored from events after all the cuts and with backgrounds.
In this context, we discuss validity, other sources of uncertainties and possibilities of improvements of our method.

We have assumed that the effects of lepton isolation and photon emissions can be evaluated and restored completely in this analysis.
Since the lepton isolation effects on the lepton energy distribution is a function of the isolation cone angle, we expect that experimental data can be extrapolated to the zero cone angle and an estimate can be obtained.
On the other hand, we can include the effect of photon emissions in the weight functions by calculating the lepton distribution with the effect.

In order to overcome the problem of the lepton cuts, we compensate for the loss using MC events.
We can also include the effects of lepton isolation in the compensating method in the same way as the effects of the lepton cuts:
by compensating for the loss caused by the lepton isolation effects.

The analysis in the previous section shows that $I(m)$ does not depend strongly on the top quark mass $m_t^{\rm c}$ of the compensated events.
This good feature is partly due to the way of determining the normalization of the compensated events.
Our strategy is to smoothly connect the lepton $p_T$ distribution, without detailed knowledge on the global shape of the distribution, which depends on PDFs.
Owing to this, the normalization of the compensated events is subject to that of the data.
If instead we utilize the total cross section to determine the normalization, we do not obtain this good feature of $I(m)$.

Quality of the fit to determine the normalization of the compensated events is a crucial factor in our method.
In this first analysis we assumed a rather simple fitting function (arbitrary quartic polynomial) and also we did not apply any correction to the $p_T$ distribution shape, after including all the cuts.
Because of this simplified analysis, we find that the quality of the present fit is not optimal.
The statistical errors in Table~\ref{tab:uncertainties} include this effect.
In a more elaborate analysis, we can estimate the (small) correction to the $p_T$ distribution caused by the cuts and also improve on the fitting function.
Alternatively it may be useful to raise the value of the lepton $p_T$ cut, since the other cuts tend to deform the lepton $p_T$ distribution more at lower $p_T$.

One may wonder if the same results can be obtained without compensating MC events.
In principle if we fit the lepton distributions to MC predictions using a multi-variate analysis, we would be able to obtain the same result.
While this is in principle possible, up to now we have not achieved to develop a pragmatic method, partly due to the complexity of such a method.
Even if this is achieved, it would be quite non-trivial to disentangle different sources of systematic uncertainties clearly.
On the other hand, an advantage of our method is that different sources of systematic uncertainties are under relatively good control.

The background estimates can be done either by a side-band method or using MC.
In the side-band method, we should take into account extrapolation errors in addition to the statistical errors.
In the case of using MC, errors of MC predictions should be considered.
The main background events after the cuts come from $t\overline{t}$ events in which $W$ bosons decay into $\mu\nu\tau\nu$, $\mu\nu e\nu$ and $\tau\nu jj$.
Since these decays can be predicted accurately, we expect that we can evaluate contributions of background events with a good accuracy.
Furthermore, the $\mu\nu\tau\nu$ and $\mu\nu e\nu$ decay modes can be included in the signal events in principle.
Besides, we can regard the decay process of the top quark where a muon is emitted via $\tau$ decay as a signal process.
In this case we include the contribution of the muon energy distribution of this process in the rest frame of the top quark into the weight functions.

We have made a separate study for the effects of $b$-tagging.
In Section~\ref{sec:cuts}, we have simulated assigning $b$-jets randomly from all the jets.
Instead we have changed the default $b$-tagging efficiency in PGS to the flat $b$-tagging efficiency $\epsilon_b=0.4$ in the region $p_T>15$\,GeV and $|\eta|<2.4$.
We have performed the same analysis as in Section~\ref{sec:results} for the case of the input top quark mass of $173$\,GeV and with a (smaller) MC event sample of $1.5\times 10^6$ events and compared the two $b$-tagging simulations.
We find that the results of the top mass reconstruction are consistent in both simulations of $b$-tagging within the MC statistical errors of about $1$\,GeV.
Nevertheless, effects of $b$-tagging would depend crucially on the details of the actual experimental conditions, and their detailed study is requisite.

With our estimates, the uncertainties associated with the factorization scale dependence in the compensated events are the main source of uncertainties in the LO analysis.
By including higher-order corrections into the production process of the top quark, we expect that these uncertainties would be reduced considerably.
In addition, including higher-order corrections into the decay process of the top quark in the weight functions, and also MC simulation if we compensate events, we can measure the $\overline{\rm MS}$ mass of the top quark with this method.
Since there are shifts due to the effects of the top quark width with the size of $0.3$-$0.4$\,GeV, it is also important to include the effects of off-shellness of top quarks.
These studies of higher-order corrections and off-shellness of top quarks are beyond the scope of this paper and left as subjects of our future works.

\section{Conclusion}
\label{sec:conclusion}

We proposed a new method to measure a theoretically well-defined top quark mass at the LHC, utilizing the ``weight function method."
This method requires the lepton energy distribution in the laboratory frame as an observable.
In an ideal limit, where the narrow-width approximation of the top quark is valid and effects of detector acceptance, event selection cuts and background contributions can be neglected, this method has a boost-invariant characteristic concerning the top quark.
Due to this characteristic, we need only the prediction of the lepton energy distribution in the rest frame of the top quark, which can be calculated in perturbative QCD.
Therefore, we can compare the observable with the perturbative QCD prediction irrespective of hadronization models and PDFs.

There are deviations from the above ideal limit.
In this paper, we concentrated on deviations due to experimental aspects, that is, the effects of detector acceptance, event selection cuts and background contributions.
Taking into account these effects, we performed a MC simulation study using $t\overline{t}$ production and lepton+jets decay channel at LO.
We found that although the effects of the lepton cuts are most serious, this difficulty can be overcome by compensating for the loss caused by the cuts using MC events.
We estimated the signal statistical error for the top quark mass to be $0.4$\,GeV corresponding to an integrated luminosity of $100$\,fb$^{-1}$ at $\sqrt{s}=14$\,TeV.
We also estimated uncertainties due to factorization scale dependence and PDF uncertainties in the compensated events, JES dependence and background statistical fluctuation.
Among these, the error due to the factorization scale dependence, which amounts to about $1.5$\,GeV, dominates.
Nevertheless, we expect that this error is reduced if we include the NLO corrections to the top quark production processes in the MC simulation.
We checked that the uncertainties from JES are suppressed, reflecting the feature of our method using the lepton observable.
Thus, our method differs qualitatively from those which use jet momenta as the primary information in the top quark mass determination.
We discussed other sources of systematic uncertainties ($b$-tagging, lepton isolation effects, quality of fits in our method, errors in background estimation, etc.) in Section~\ref{sec:discussion}.
In conclusion, we estimate that various systematic uncertainties can be sufficiently tamed in our method.
It is, however, imperative toward a realistic top quark mass determination to incorporate in the analysis the details of the actual experimental conditions at the LHC.
In this respect, an analysis in collaboration with experimentalists would be desirable in the future.

Taking into account theoretical corrections to the ideal limit will be subjects of our future works.
Important corrections are the NLO and NNLO corrections to the top quark decay processes in perturbative QCD~\cite{Jezabek:1988ja, Gao:2012ja, Brucherseifer:2013iv} and the effects of the off-shellness of the top quark.
As a consequence of the boost-invariant nature, this method has the advantage that only the higher-order QCD corrections concerning the decay process of the top quark are primarily required.
By including these corrections in weight functions, the $\overline{\rm MS}$ mass of the top quark can be determined.
Since we use MC events in the compensating method which we devised to overcome the problem of the lepton cuts, we should include also in the MC simulation the corrections to the leptonic decay.

We point out that this method can be applied to the di-leptonic channel as well.
Although this channel has a smaller cross section than the lepton+jets channel, it is clean and (in principle) a lower lepton $p_T$ cut can be applied.
Thus, the di-leptonic channel is also worth investigating.

At present a precise measurement of the top quark $\overline{\rm MS}$ mass at the LHC is highly demanded but regarded as quite challenging.
We hope that our present study provides a useful basis in this direction.

We are grateful to H. Kawai and K. Fujii for valuable discussion and comments.
The works of S.K. and Y. Sumino, respectively, were supported by Grant-in-Aid for JSPS Fellows under the program number 24$\cdot$3439 and by the Japanese Ministry of Education, Culture, Sports, Science and Technology by Grant-in-Aid for Scientific Research under the program number (C) 80260412.
The work of H.Y. was supported in part by Grant-in-Aid for Scientific Research, No.\,24340046 and the Sasakawa Scientific Research Grant from the Japan Science Society.


\end{document}